\documentclass[12pt,preprint]{aastex}

\usepackage{natbib}

\shorttitle{Infrared Emission of Abell~58}

\shortauthors{Koller \& KIMESWENGER}

\received{Jan. 19, 2001} \revised{} \accepted{}

\ccc{}
\cpright{AAS}{2001}

\begin{document}


\title{THE ANOMALOUS INFRARED EMISSION OF ABELL~58}

\author{Josef Koller\altaffilmark{1}}
\affil{Department of Physics \& Astronomy, MS-108, Rice
University, Houston, TX 77005\\
Theoretical Astrophysics, T-6, MS B288, Los Alamos National
Laboratory, NM 87545} \altaffiltext{1}{Email: {\tt
josef@rice.edu}}

\and

\author{Stefan Kimeswenger\altaffilmark{2}}
\affil{Institute of Astrophysics, University of Innsbruck,
A-6020 Innsbruck, Austria} \altaffiltext{2}{Email: {\tt
stefan.kimeswenger@uibk.ac.at}}

\begin{abstract}

We present a new model to explain the excess in mid and near
infrared emission of the central, hydrogen poor dust knot in the
planetary nebula (PN) Abell~58. Current models disagree with ISO
measurement because they apply an average grain size and
equilibrium conditions only. We investigate grain size
distributions and temperature fluctuations affecting infrared
emission using a new radiative transfer code and discuss in detail
the conditions requiring an extension of the classical
description. The peculiar infrared emission of V605~Aql, the
central dust knot in Abell~58, has been modeled with our code.
V605~Aql is of special interest as it is one of only three stars
ever observed to move from the evolutionary track of a central PN
star back to the post-AGB state.
\end{abstract}

\keywords{ISM: dust, planetary nebulae: individual (V605~AQL),
stars: circumstellar matter, methods: numerical} \sloppy
\section{INTRODUCTION}

The first evidence of a significant discrepancy between
measurements and predictions of standard models for infrared (IR)
emission from dust was noticed by \citet{Andriesse78}. He found
that IR emission of the \ion{H}{2} region M17 and the diffuse
galactic emission is at 11~\micron\ and 20~\micron\ considerably
higher than standard predictions \citep{Mathis83, Puget89}.
\citet{Andriesse78} pointed out that spatial temperature
variations cannot solve the conflict between model predictions and
measurements, but temperature fluctuations of the dust grains
themselves could. In general, small dust particles have a lower
heat capacity due to their smaller size. Therefore, particles with
radii smaller than 20~\AA\ to 50~\AA\ follow temperature
fluctuations, but this range strongly depends on the radiation
field. Moreover, if the grain has time to cool down, i.e. no
continuous photon impact and heating, the emitted IR photons of a
sample of dust grains have a spectral energy distribution (SED)
which differs from that in the equilibrium case. The temperature
of such grains can be modeled by means of a temperature
distribution function $P_{a} (T)$. Various groups have studied
temperature fluctuations of small grains and proposed numerical
methods for the calculation \citep{Gail75, Purcell76, Draine85,
Desert86, Dwek86}. However, a more powerful technique was
introduced by \citet{Guhathakurta89} to find the temperature
distribution and the associated infrared emission. This method can
be easily implemented into a computer code and is reasonably fast.
A further description of the technique can be found in
\citet{Borkowski94, Siebenmorgen92}, and an advanced extension of
the method was published by \citet{Manske98}.

We discuss the emission of real grain size distributions for
thermal equilibrium and non-equilibrium conditions in the
following section. We applied our new radiative transfer code to
the central dust knot V605~Aql in Abell~58 (a result of a
late-helium flash in 1919). A proper model of the current state of
the circumstellar shell, completely shielding the new central
star, is important to understand the physical process of dust
formation and of the mass ejection history. This object
illustrates stellar evolution in real time. Our model of the
circumstellar shell around the central star does not only use the
dust component, rather it combines gas and dust to obtain a more
accurate description of V605~Aql.

\section{THERMAL EQUILIBRIUM AND NON-EQUILIBRIUM}

Temperature is one of the most important physical parameters of a
dust grain. In this work only heating by photons is considered.
The calculation of  the temperature of a dust grain is, under the
assumption of thermal equilibrium conditions, usually
straight-forward, i.e., the energy absorbed from the ambient
radiation field is precisely matched by the energy emitted, and
thus an equilibrium temperature is attained:
\begin{equation}
\begin{array}{l}
c\int \limits_{0}^{\infty} u_{\lambda}Q_{abs}(a,\lambda)\ d\lambda =\\
\qquad\qquad4\pi \int \limits_0^{\infty} Q_{abs}(a,\lambda)
B_{\lambda} (T_{d})\ d\lambda
\\
\end{array}
\end{equation}
where $u_\lambda$ is the density of the heating radiation field.
$Q_{abs}(a,\lambda)$ are the efficiency factors for absorption for
a dust grain with radius $a$ at different wavelengths $\lambda$,
and $B_\lambda (T_{d})$ is the blackbody emission of a grain at
temperature $T_{d}$. In the case of thermal equilibrium it is
sufficient to to solve the equation above, but a size distribution
of dust grains has to be included.

Laboratory experiments and theoretical considerations suggest that
the mass distribution of dust particles is described as
\begin{equation}
n(m)  dm \propto m^{-p}  dm
\end{equation}
where $p$ is a constant and $n(m)$ is the number of grains in the
range from $m$ to $m+dm$. \citet{Mathis77}, hereafter MRN, applied
this relation to extinction measurements of the interstellar
medium and found a power law of $p=3.5$ for the general
interstellar medium (ISM). Infrared objects have a much higher
excess in mid-infrared and near-infrared (MN-IR) than simple dust
models predict, but applying a grain size distribution can lead to
a better agreement between measurements and models. However, if
the dusty material is situated in a dilute radiation field, then
the assumption of thermal equilibrium is not valid. The dust
temperature changes due to the stochastic process of grain heating
and has to be treated in a more sophisticated way as described
below.

Considering a spherical dust particle of radius $a$ and with a
heat capacity per unit volume $C(T_{d})$, the internal energy of a
dust particle is calculated using
\begin{equation}
U(a,T_{d}) = \frac{4\pi a^{3}}{3}\int_{0}^{T_{d}} C(T')\ dT'.
\end{equation}
If an incident photon deposits the energy $E$ on the dust
particle, the energy will be distributed almost instantly and will
raise the temperature of the grain by an amount of $\delta T_{d}$
according to $E=U(a,T_{d}+\delta T_{d}) - U(a,T_{d})$. If the dust
grain is sufficiently small and thus  $E\gg U(a,T_{d})$, the dust
particle will experience a strong rise in temperature. For larger
dust grains or high photon densities, the energy provided by a
single photon is negligible compared to the internal energy of the
grain. In this case the dust grain will fluctuate with a small
amount around the equilibrium temperature, i.e., the temperature
distribution function of the dust grain is approximately a small
Gaussian centered on $T_{d}$ which can be written as a delta
distribution for very small fluctuations.

Temperature fluctuations have to be taken into account, if: (1)
The density of the radiation field is low,  (2) the radiation
field contains high-energy photons providing the grain with a
large amount of energy compared to the internal energy, (3) the
grain size is small ($a\,<\,100$~\AA) and, therefore, internal
energy is low. These circumstances are obviously met for small
grains in the ISM or for PNe, because the radiation field is weak
but contains many high-energy photons.

\section{THE HYDROGEN--POOR KNOT V605~AQL}
        \subsection{Introduction to V605~Aql}

Some central stars of planetary nebulae (PNe) experience a final
thermal pulse \citep{Iben83} after starting the descent along the
cooling track towards white dwarfs. Theoretical calculations show
that during such a pulse the remaining hydrogen is completely
incorporated into the helium-burning shell. The object then
briefly expands to the dimension of an AGB star and proceeds with
helium burning. It follows nearly the same path in the
Hertzsprung-Russell diagram (HRD) as during the hydrogen burning
when the initial excitation of the nebula occurred. The evolution
is very fast, and \citet{Iben83} suggested that it would be too
short to be observable. However,  at least three stars have been
found during their rapid brightening caused by a final helium
flash. (1) Sakurai's Object or V4334 Sgr was caught during a
nova-like outburst in 1996. (2) FG Sagittae brightened in the year
1894. (3) V605~Aql was discovered in 1919 and was mistaken to be a
slow nova. Since these objects evolve very quickly, especially
Sakurai's object, on time-scales from weeks to years, they provide
the rare opportunity for studying stellar evolution in real time.

The evolution of the late helium flash object V605~Aql
(IRAS~19158+0141) is reviewed in the paper by \citet{Clayton97}.
Starting in August 1919 the object brightened within two years
from $m_{pg}\,\approx\,15^{\rm m}$ to
$m_{pg}\,\approx\,10\,\,.\!\!\!^{\rm m} 2$. Its peculiarity was
noted in a spectrum in 1921, but after a few phases of fading and
brightening it got too weak, and therefore nothing can be found in
the literature until 1971 \citep{Ford71}. At that time V605~Aql
was recognized as a star/nebulosity (diameter $\sim 1^{\prime
\prime }$) in the geometric center of the PN Abell~58
\citep{Abell66}, which has a size of $44^{\prime \prime } \times
36^{\prime \prime }$. The spectrum of the outer nebula Abell~58 is
typical for a PN. However, it has a weaker H$_{\alpha }$ than
[\ion{N}{2}] emission. \citet{Seitter87} took a spectrum of the
central knot V605~Aql after \citet{Pottasch86} suggested that it
is a hydrogen poor object. Seitter derived physical conditions for
the inner and outer nebula from the lines [\ion{O}{3}] (4959~\AA ,
5007~\AA ) along with [\ion{N}{2}] (6548~\AA , 6584~\AA ) and
[\ion{O}{1}] (6300~\AA , 6363~\AA ). However, these values are
quite uncertain and yield contradictory results from different
spectra due to high noise. The abundance of elements can only be
estimated because of the weakness of diagnostic lines. Recent
determinations for both, Abell~58 and V605~Aql are made by
\cite{Guerrero96}.

\citet{Seitter87} found in her spectra that V605~Aql shows a
\ion{C}{4} feature (5800 \AA ), which suggests that the central
star is of Wolf-Rayet [WC] type, but it was not possible to
clearly differentiate this feature from non-stellar lines.
However, in \citet{Guerrero96} it is clearly identified. The
correction for dust absorption is ''merely a guesswork''
\citep{Seitter87}. With $T=100,000$~K and $A_{V}=4^{\rm m}$ at a
distance of 3.5~kpc, the obtained luminosity is 300~L$_{\odot}$
but this would place the object in the HRD at a position which is
reached 1000 years after the late helium flash for a 0.5~M$_{\odot
}$ star. Compared to the reported nova event in 1919 this would be
one order of magnitude too large. Moreover, the integration over
the infrared spectrum at $3.5$ kpc leads to a luminosity of about
2500~L$_{\odot }$, which is still too low to account for the
recent event. Therefore, we adopt a distance of 5 kpc to place
V605~Aql at a proper position (5800~L$_{\odot }$) in the HRD.

\citet{Seitter87} reports a radial expansion velocity of the inner
knot of 60~km~s$^{-1}$. In contrast, \citet{Pollacco92} find with
high resolution and high S/N spectra a value of about
$100$~km~s$^{-1}$. The outflow of the central knot in Abell~58 is
completely hydrogen depleted, and the ISO spectra taken by
\citet{Harrington98} show no PAH feature, as is expected for a
H-poor carbon rich outflow.

Observations with ISOCAM yield a bright MN-IR emission from the
central knot in Abell~58. \citet{Kimeswenger98} concluded that a
simple dust emission model cannot explain the SED. The  ISO
observations, which we use here (Table \ref{ISOvalues}), are
recalibrated with ISO data reduction software OLP V.5
\citep{Siebenmorgen99}. Note that the extreme 1 to 3~$\mu$m excess
reported in \cite{Kimeswenger98} is caused by a misidentification
in the NIR photometry in the past (see \citealt{Kimeswenger00} for
further details).

\placetable{ISOvalues}

       \subsection{The Model for V605~Aql}

\begin{figure}
\plotone{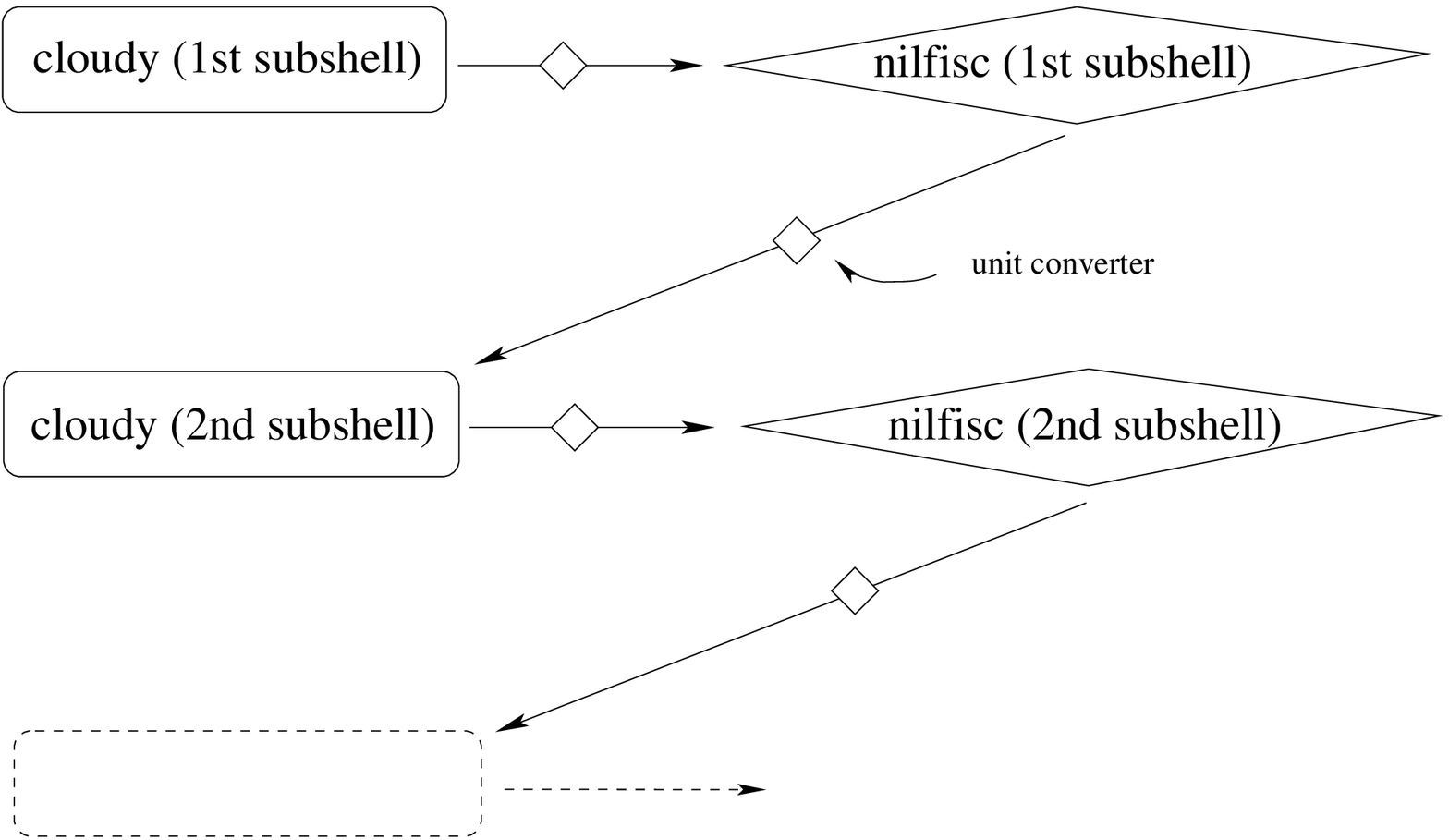} \caption{Scheme for the calculation sequence. The
first subshell of gas is computed with CLOUDY which serves the
output radiation to NILFISC. NILFISC calculates the extinction and
emission from the dusty component and delivers the total output
radiation of the first subshell (gas and dust) to CLOUDY and so
forth.} \label{scheme}
\end{figure}

We linked our numerical code NILFISC (Near Infrared Light From
Interstellar Scattering Code) to the photoionization code CLOUDY
\citep{Ferland98} to combine gas and dust models. This was implied
by using small shells of gas (CLOUDY) and dust (NILFISC) providing
each other with the input radiation (Fig. \ref{scheme}). CLOUDY
calculates the gas continuum and emission lines of the first shell
and serves the radiation as an input for NILFISC. NILFISC then
adds extinction and dust emission to this shell and provides the
radiation for the next gas shell calculated again by CLOUDY and so
forth. The number of sub-shells required for this calculation
technique were extensively tested to (a) optimize the CPU time and
(b) to minimize numerical difficulties. We describe numerical and
stability tests for various types of sources in \citet{Koller99,
Koller2001}. A more extensive and detailed version can be found in
\citet{KollerMSc00} which is obtainable as a postscript version
from the authors upon request.

\placefigure{scheme}

The dust shell model in V605~Aql utilizes a decreasing dust
density $\propto 1/r^2$, which is consistent with the model of a
steady outflow at a constant velocity during a short nucleation
period \citep{Rowan83}. The outflow is carbonaceous and contains
no silicate or PAH signatures \citep{Harrington98}. Therefore, we
applied the optical properties of carbon grains \citep{Laor93} to
model this object. The value of $A_{V}=7^{\rm m}$ is in the range
of preceding extinction evaluations. We chose $A_{V}$ according to
the literature and configured our dust shell density for this
value finding a gas to dust ratio of 10/1 in an almost H-less
outflow. On the basis of the youth of the object, a filling factor
near unity can be assumed as has been already successfully applied
to other very young optical thick shells \citep{Siebenmorgen94}.
We used a blackbody central star with a temperature of $100,000$~K
and a luminosity $5800$~L$_{\odot}$ as the heating source for the
dust shell. These values are supported by evolutionary track
models. At a smaller distance the required luminosity would be too
low compared to theoretical predictions \citep{Bloecker95b}. At a
larger distance the old PN, Abell~58, would become too large
compared to its surface brightness. By fixing the distance to 5
kpc, a lower boundary for the total luminosity can be found by
integration of the SED defined by the ISOCAM and IRAS
measurements.

The inner radius and the grain size distribution inside the dust
shell are left as free parameters. However, the power law of the
grain size distribution is assumed to be constant throughout the
rather small shell (5$\times 10^{13}$ m). A limit of the outer
radius can be set to a maximum of 2\arcsec\ \citep{Geckeler99}.
This is also in agreement with optical measurements by
\citet{Guerrero96}. The parameters for the shell size are
supported by velocity determinations by \citet{Pollacco92}
providing $v_{\exp }\approx 100$~km~s$^{-1}$ for the inner knot.
However, the line profile wings have not been taken into account,
and according to \cite{Schoenberner99} such expansion estimations
are too low by 20-30\%, yielding $v_{\exp }\approx
120$~km~s$^{-1}$. Assuming a constant $v_{\exp }$, this value
leads to a shell radius of $2.6\times 10^{14}$~m after 70 years,
which was the age of the object at the time of measurements by
\cite{Pollacco92}. The dust shell size, the total extinction, and
the assumed density profile provide us with the total dust mass.

The H-poor nature of V605~Aql was taken into account by using a
virtually H-less gas mixture in CLOUDY. We chose H/He $\approx
10^{-6}$ below the solar value to simulate the H under-abundance
(other elements in respect to He). This approach removes the
912~\AA\ absorption edge of hydrogen, and the He ionization at 54
eV becomes dominant for the gas absorption. This leaves more UV
photons to be absorbed by small dust particles, which have major
absorption efficiencies at 750 \AA\ and 2200 \AA\ (famous
extinction interstellar extinction bump). Usually the first
feature is not very efficient because hydrogen Lyman absorption
between 13 eV and 54 eV has already removed most of the absorbable
photons.

\section{RESULTS}

\begin{figure}
\plotone{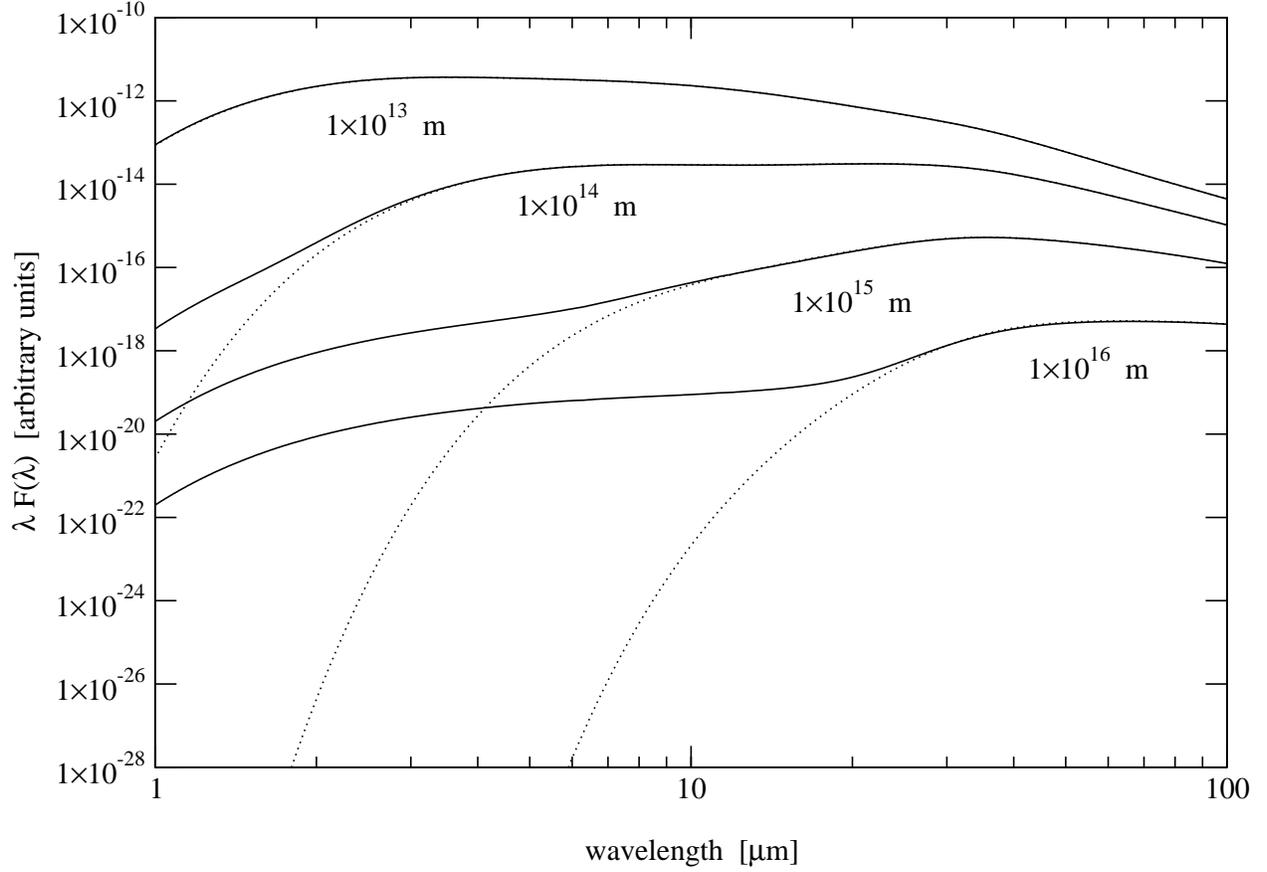} \caption{The emission of a thin, pure dust shell
with temperature fluctuations (solid line) and the emission from
equilibrium calculations (dotted line) at various distances to the
central star. Illumination and dust parameters of the model are
described in Table \ref{parameter_tab}. Note the deviations
between the two methods. } \label{fluctus_equi}
\end{figure}

Figure \ref{fluctus_equi} shows the emission of a pure dust shell
for various distances from the central star. Both methods, with
temperature fluctuations and equilibrium temperatures, were used
to investigate at which distance they start to differ.
In the case of such a high luminosity object ($L_{\ast }\sim 6000$ L$%
_{\odot },$ $T\sim 100,000\;$K) and with dust grains at a distance of 10$%
^{13}$~m, temperature fluctuation generally do not arise. If the
grains are at a distance of 10$^{14}$~m, the two methods start to
show a discrepancy at $\lambda \lesssim 3$ $\mu $m. For grains at
larger distances to the central star the discrepancy rises quickly
up to larger wavelengths. For grains at 0.3 pc (10$^{16}$~ m) the
two methods differ already for wavelengths $\lambda \lesssim
11$~$\mu $m. Note that this is only true for a grain size
distribution.

\placefigure{fluctus_equi}

\begin{figure}
\plotone{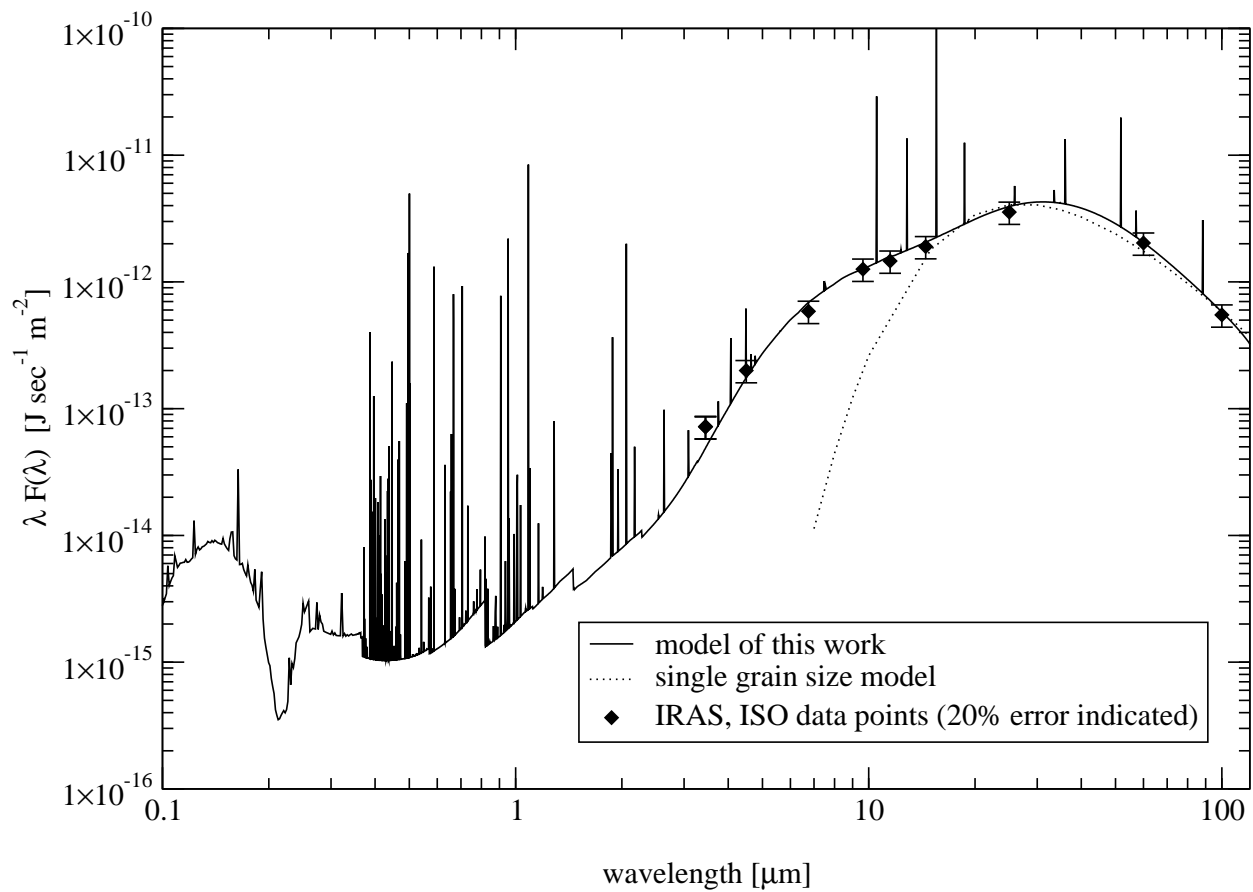} \caption{The model for V605~Aql using the
emission from fluctuating dust grains. It has been calculated with
the parameters given in Table \ref{parameter_tab}. Diamonds
represent the observations (Table \ref{ISOvalues}) with 20 percent
error bars. The dotted line is the single grain size model after
\cite{Pollacco92}. Emission lines and absorption bands are due to
the gaseous component in our model. } \label{final_result}
\end{figure}

We derived a dust shell radius of $2.5-3.0\times 10^{14}$ m for
V605~Aql, which means that we are dealing with a ''borderline''
object where temperature fluctuations start to get important
(Figure \ref{fluctus_equi}), although the ionization edges of the
gas and internal extinction of the object further dilutes the UV
radiation field. We conclude that V605~Aql is actually on the
lower boundary where temperature fluctuations have to be
considered. In V605~Aql the emission by temperature fluctuations
may be approximated in first order by applying the method of
equilibrium temperature computations, but a single grain size
model is not a proper approximation (see Figure
\ref{final_result}). This Figure also shows the gas emission
taking over at wavelengths about 1~$\mu$m. The strong but narrow
emission lines, visible in the model, do not significantly
contribute to the wide band flux. Table \ref{parameter_tab}
provides the parameters we used for our model.

\placetable{parameter_tab}
\placefigure{final_result}

\begin{figure}
\plotone{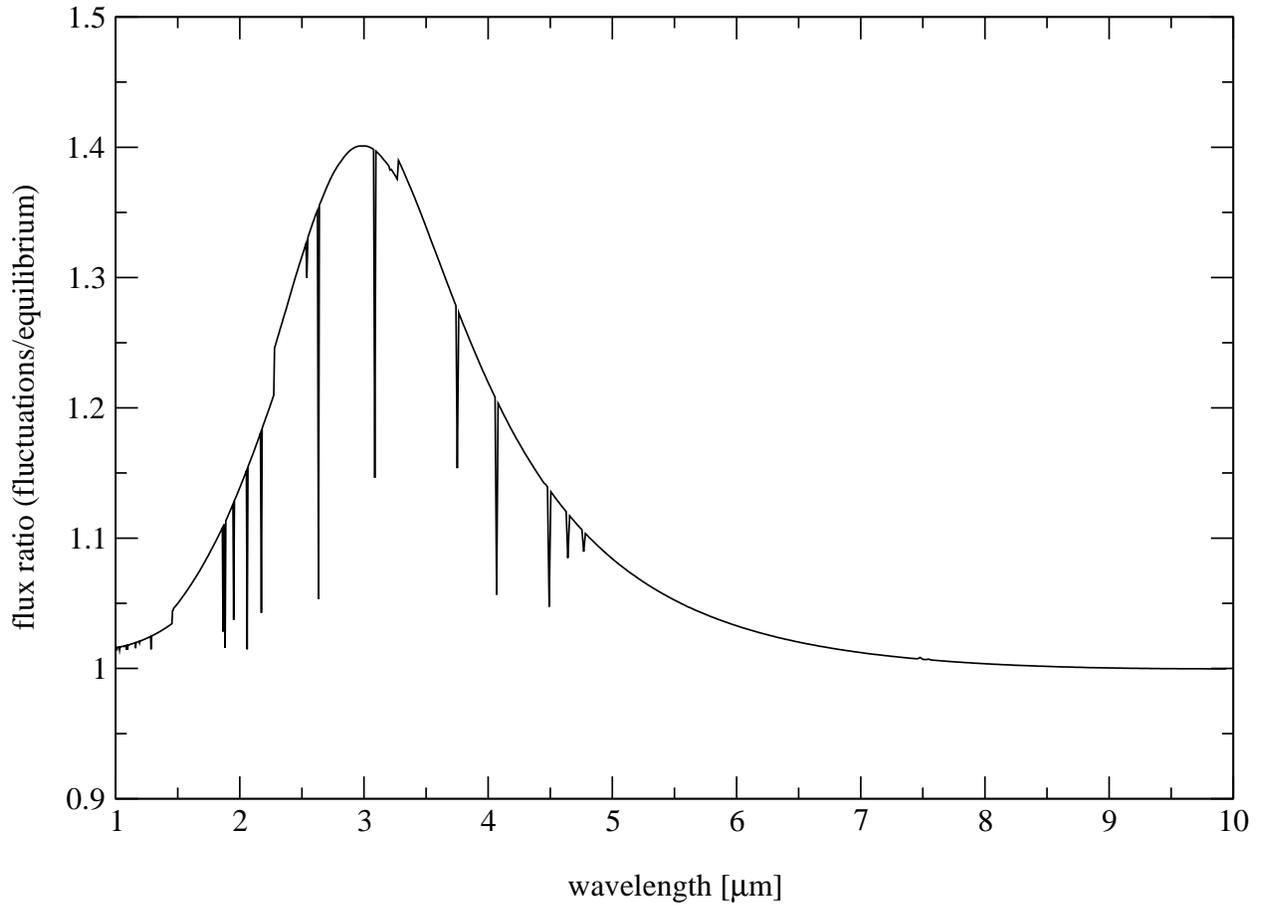} \caption{The ratio of the model using temperature
fluctuations compared to using the emission by grains in
equilibrium as a function of wavelength. } \label{m23_24}
\end{figure}

In Figure \ref{m23_24} we compare two different models (with and
without temperature fluctuations). V605~Aql is affected mainly in
the wavelength range of 2~$\mu$m~$\le\,\,\lambda\,\,\le$~5~$\mu$m.
This is on one hand due to the fact that the shell radius is
rather small and on the other hand, the hydrogen, usually
depleting the UV radiation below 91.2 nm, is strongly
under-abundant.

\placefigure{m23_24}

Compared to the dust mass predicted by evolutionary model
computations \citep {Bloecker95}, the dust mass in our model is
rather high. However, the model assumes a spherically symmetric
dust shell requiring a copious amount of dust. \cite {Pollacco92}
did not exclude that the ejection could be in the form of a
symmetric shell. It could also be a bipolar outflow of which we
only see the blue shifted side, as also indicated in
\citet{Guerrero96}. Non-spherical shell models are planned to
reduce this discrepancy. The current version, although now
applying spherical symmetry, was written highly modular. Using
small plane parallel slabs, will allow us to extend the code to 2D
and 3D geometry in nearby future.

\section{CONCLUSION}

The IR emission caused by the hot components of dust grains
undergoing temperature fluctuations due to single-photon heating
events is observed in a variety of regions: in the interstellar
medium, in H II regions, reflection nebulae, planetary nebulae,
interstellar medium, cirrus, and at the edges of dense molecular
clouds. The dependence of the distribution of this emission upon
the spectral energy distribution shows that UV photons from very
hot stars dominate this mechanism.

The results from the code NILFISC, developed for the calculations
presented here, show that introducing a real grain size
distribution improves the situation significantly and that a
equilibrium temperature model with a single average grain size is
not sufficient. However, the use of the temperature fluctuation
technique for small dust grains is not necessary in dense
radiation fields and much computing time can be saved if the
sophisticated method with stochastic heating is not required. Our
code evaluates the situation for every grain size individually.

The model of the born-again core V605~Aql of the PN Abell~58 is
able to describe the ISO and IRAS observations. Although the dust
shell is small and the heating source very luminous, the
application of a grain size distribution but only with equilibrium
calculations yield in a discrepancy of about 40 percent at short
IR wavelengths. V605~Aql is a borderline case: at lower radii,
higher luminosities, or a cooler illuminating source, temperature
fluctuations do not have to be taken into account.

In contrary to previous work \citep{Seitter87, Guerrero96,
Clayton97}, we had to adopt a larger distance (5 kpc) and a higher
luminosity (5800 L$_{\odot}$ for a satisfactory fit of our model
to observations. The higher luminosity is in much better agreement
to evolutionary models \citep{Bloecker95, Bloecker95b} than the
derived luminosity of e.g. 300 L$_{\odot}$ in \cite{Seitter87}. To
keep consistency with the observed flux, the higher luminosity
requires a larger distance (5~kpc) than the usually cited value of
3.5~kpc \citep{Clayton97}.

Because most planetary nebulae are larger and have hot central
stars, the technique with temperature fluctuations ought to be
applied to most of them.

\acknowledgments This work has been supported by the Austrian {\it
Fond zur F\"orderung der wissenschaftlichen Forschung}, project
number P11675-AST. We would like to thank P.~Hartigan and the
referee for valuable suggestions and comments.

\clearpage
%
%

\clearpage

\clearpage

\clearpage

\clearpage

%

\begin{deluxetable}{llll}
\tabletypesize{\footnotesize}
\tablecaption{IR Measurements Used in Modeling\label{ISOvalues}}
\tablewidth{0pt}
\tablehead{
\colhead{Instrument} & \colhead{$\lambda$ [$\mu$m]} &
\colhead{F$_{\nu}$ [Jy]} & \colhead{$\lambda F_{\lambda}$ [W
m$^{-2}$]} } \startdata
L \tablenotemark{a} & 3.45 & $0.083$ & $7.21\times 10^{-14}$ \\
ISO LW1 & 4.50 & $0.3$ & $2.00\times 10^{-13}$ \\
ISO LW5 & 6.75 & $1.3$ & $5.87\times 10^{-13}$ \\
ISO LW7 & 9.63 & $4.1$ & $1.26\times 10^{-12}$ \\
ISO LW8 & 11.50 & $5.6$ & $1.46\times 10^{-12}$ \\
ISO LW9 & 14.5 & $9.2$ & $1.90\times 10^{-12}$ \\
IRAS LW7 & 25 & $29.5$ & $3.54\times 10^{-12}$ \\
IRAS LW10 & 60 & $40.6$ & $2.03\times 10^{-12}$ \\
IRAS LW13 & 100 & $18.3$ & $5.49\times 10^{-13}$
\enddata
\tablenotetext{a}{The L band value has been taken from \cite{Veen89}}
\end{deluxetable}

\clearpage

\begin{deluxetable}{cl}
\tablewidth{0pt} \tablehead{ \colhead{Parameter} & \colhead{Value}
} \tablecaption{Model Parameters for V605~Aql}

\startdata

Gas shell &  \\
\multicolumn{1}{l}{Total gas mass} & $7.6\times 10^{-2}$ $M_{\odot }$ \\
\multicolumn{1}{l}{Hydrogen abundance} & H/He $\approx 10^{-6}$ solar \\
\multicolumn{1}{l}{Heavy Elements} & He/Y solar \\
&  \\
Dust shell &  \\
\multicolumn{1}{l}{Inner radius} & $r_{in}=2.5\times 10^{14}$ m \\
\multicolumn{1}{l}{Outer radius} & $r_{out}=3.0\times 10^{14}$ m \\
\multicolumn{1}{l}{Total dust mass} & $8.0\times 10^{-3}\;M_{\odot }$ \\
\multicolumn{1}{l}{Visual extinction} & $A_{V}=7$ mag \\
&  \\
Dust material &  \\
\multicolumn{1}{l}{Carbonaceous grains} &  \\
\multicolumn{1}{l}{Grain size distribution (MRN)} & $a_{\min }=6\;$\AA ,$%
\;a_{\max }=4.55\;\mu $m$,\;p=3.1$ \\
&  \\
Central star &  \\
\multicolumn{1}{l}{Luminosity} & $L=5800\;L_{\odot }$ \\
\multicolumn{1}{l}{Temperature} & $T=100,000\;$K black body \\
&  \\
\multicolumn{1}{l}{Adopted distance to Abell~58} & $d=5\;$kpc
\enddata
\label{parameter_tab}
\end{deluxetable}

\end{document}